\documentclass[aps,prx,reprint,superscriptaddress,floatfix,longbibliography]{revtex4-2}
\usepackage[T1]{fontenc}
\usepackage{lmodern}
\usepackage[utf8]{inputenc}
\usepackage{amsmath,amssymb,bm}
\usepackage{graphicx}
\graphicspath{{figures/}}
\usepackage{booktabs}
\usepackage{xcolor}
\definecolor{ArticleLinkBlue}{HTML}{2E3092}
\usepackage{hyperref}
\hypersetup{colorlinks=true,
  linkcolor=ArticleLinkBlue,
  citecolor=ArticleLinkBlue,
  urlcolor=ArticleLinkBlue,
  filecolor=ArticleLinkBlue,
  pdftitle={Local Dephasing Sustains Continuous Time Crystals in a Spin-1/2 Ensemble},
  pdfauthor={Zhen-huan Yang and Dan-Bo Zhang}}

\newcommand{\Tr}{\operatorname{Tr}}
\newcommand{\stead}{\mathrm{ss}}
\newcommand{\osc}{\mathrm{osc}}

\newcommand{\ii}{\mathrm{i}}
\newcommand{\dd}{\mathrm{d}}
\newcommand{\avg}[1]{\left\langle #1\right\rangle}

\begin{document}
\title{Local Dephasing Sustains Continuous Time Crystals in a Spin-1/2 Ensemble}
\author{Zhen-huan Yang}
\affiliation{Key Laboratory of Atomic and Subatomic Structure and Quantum Control (Ministry of Education),  Guangdong Basic Research Center of Excellence for Structure and Fundamental Interactions of Matter, and  School of Physics, South China Normal University, Guangzhou 510006, China}

\author{Dan-Bo Zhang}
\email{dbzhang@m.scnu.edu.cn}
\affiliation{Key Laboratory of Atomic and Subatomic Structure and Quantum Control (Ministry of Education),  Guangdong Basic Research Center of Excellence for Structure and Fundamental Interactions of Matter, and  School of Physics, South China Normal University, Guangzhou 510006, China}
\affiliation{Guangdong Provincial Key Laboratory of Quantum Engineering and Quantum Materials,  Guangdong-Hong Kong Joint Laboratory of Quantum Matter, and Frontier Research Institute for Physics,\\  South China Normal University, Guangzhou 510006, China}

\date{October 20, 2026}
\begin{abstract}
Local pure dephasing suppresses single-spin coherence, yet we show that it sustains continuous time-crystalline order in a fully connected ferromagnetic spin-$1/2$ ensemble with local decay.
An exact operator identity reveals the mechanism: dephasing raises the interaction energy, while energy relaxation on a finite timescale converts this charging into delayed feedback.
Within homogeneous mean-field theory, we prove that dephasing is necessary for periodic motion in this model and that feedback exactly balances transverse damping over each cycle.
Beyond mean field, the identity fixes the relative amplitude and phase of energy and longitudinal magnetization in oscillatory Liouvillian modes.
Liouvillian spectra and stationary two-time correlations characterize temporal order through decay rates that decrease algebraically with system size, frequencies approaching nonzero mean-field values, and comparable per-spin oscillatory weights.
Semiclassical trajectories further reveal recurrent switching between a limit-cycle region and two symmetry-related stationary regions: fluctuations interrupt and reactivate collective oscillations, and oscillatory episodes provide an intermediate channel between stationary regions.
These results establish local dephasing as a resource for autonomous collective motion and show how the resulting temporal order can participate in metastable dynamics.

\end{abstract}
\maketitle
\raggedbottom

\section{Introduction}
\label{sec:introduction}

\begin{figure*}[t]
	\centering
	\includegraphics[width=\textwidth]{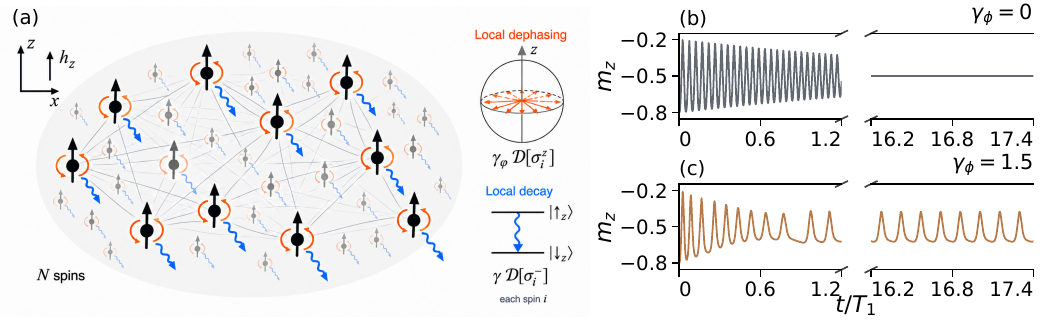}
	\caption{Dephasing-controlled collective oscillations.
	    (a) Fully connected ferromagnetic ensemble with independent local spin decay and pure dephasing.
	    (b,c) Longitudinal magnetization at $\gamma_\phi=0$ and $1.5$, respectively, from the same initial Bloch vector $\boldsymbol m(0)\approx(0.39,0.03,-0.80)$.
	    Both use $T_1=1/\Gamma_1=1$, $J\approx105.4$, and $h_z\approx105.4$; rates are in units of $T_1^{-1}$.
	    Broken axes show $t/T_1\in[0,1.2]$ and $[16.2,17.4]$, with identical vertical scales.}
	\label{fig:phenomenon}
\end{figure*}

Continuous time crystals extend spontaneous ordering to persistent collective motion~\cite{Wilczek2012,Watanabe2015,Iemini2018,Booker2020,Carollo2022}. Unlike periodically driven discrete time crystals~\cite{Else2016,Khemani2016,Yao2017,Zhang2017,Gong2018,Lazarides2020,RieraCampeny2020,Kessler2021}, their oscillations arise under time-independent dynamics. Dissipation can support this order, extending its constructive role in quantum-state engineering~\cite{Diehl2008,Verstraete2009}. Collective dissipation couples constituents through shared environmental channels~\cite{Tucker2018,Piccitto2021,Lee2014,Shammah2018}; with independent local dissipation, interactions can instead organize their motion. Such interaction-based oscillations have been explored in spin and atomic ensembles~\cite{Lee2011,Qian2012,Chan2015,Yang2025,Wang2025,Russo2025}, photonic systems~\cite{Jin2014,Seibold2020,Alaeian2021,Bakker2022}, and coupled ensembles~\cite{Nadolny2023,Solanki2024,Yang2026}. Experimental platforms include atom--cavity systems~\cite{Kongkhambut2022,Dreon2022,Cosme2025}, Rydberg gases~\cite{Wadenpfuhl2023,Wu2024,Liu2025,Jiao2025}, rare-earth-doped solids~\cite{Chen2023}, and spin ensembles~\cite{Greilich2024,Huang2025}.

Many routes to dissipative collective oscillations employ population-changing processes, such as spontaneous decay or incoherent pumping, that directly redistribute local occupations. Pure dephasing presents a different possibility. It is ubiquitous in realistic spin ensembles and is generally regarded as unwanted decoherence: it suppresses coherence in its preferred basis without directly changing local populations or providing an obvious excitation channel. Although dephasing can strongly modify interacting nonequilibrium dynamics~\cite{Poletti2012,Cai2013,Kirton2017}, its potential to sustain autonomous temporal order remains unclear. Can local pure dephasing itself become essential for continuous time-crystalline order?

Here we demonstrate how local pure dephasing sustains collective temporal order in a fully connected ferromagnetic spin-$1/2$ ensemble with independent local decay. Our results comprise four main contributions.
\textit{(i) Dephasing-driven oscillations and their mechanism.} An exact operator identity separates decay-driven relaxation of a shifted energy observable from a dephasing-induced source proportional to transverse pair correlations. Within homogeneous mean-field theory, we prove that periodic collective oscillations require pure dephasing when local decay is present. Energy charging alone, however, does not sustain motion: the finite energy-relaxation time generates delayed feedback on the transverse magnetization. We show that the cycle-integrated work of this feedback exactly compensates transverse damping along a periodic orbit.
\textit{(ii) Quantum signatures of time-crystalline order.} Beyond mean field, the same operator identity determines the relative amplitude and phase of energy and longitudinal magnetization in quantum oscillatory Liouvillian modes. Liouvillian spectra and stationary two-time correlations reveal decay rates decreasing algebraically with system size, frequencies approaching nonzero mean-field values, and appreciable per-spin collective oscillatory weights. These results connect the microscopic energy mechanism to collective quantum signatures of continuous time-crystalline order.
\textit{(iii) Intertwined oscillations and metastability.} Semiclassical stochastic trajectories reveal that the limit cycle coexists with two symmetry-related stable fixed points. Intrinsic fluctuations repeatedly interrupt collective oscillations with periods of residence near either fixed point and subsequently reactivate oscillatory motion. The oscillatory region also participates in switching between the two stationary regions, while occupation-time statistics characterize the resulting intermittency.
\textit{(iv) Observability and robustness.} Semiclassical photon-counting simulations show how the longitudinal oscillations can be resolved in detected spectra, and parameter-noise tests assess their robustness. Thus, a simple and widely used interacting-spin model with only time-independent local dissipative channels displays both sustained temporal order and a rich organization of dynamical attractors.

The paper is organized as follows. Section~\ref{sec:model} introduces the model and mean-field dynamics. Sections~\ref{sec:mechanism} and \ref{sec:quantum} establish the energy-feedback mechanism and its quantum signatures. Section~\ref{sec:metastability} examines metastable intermittency, Sec.~\ref{sec:observability} addresses observability and robustness, and Sec.~\ref{sec:discussion} discusses the implications and concludes.

\section{Collective oscillations enabled by local dephasing}
\label{sec:model}

\subsection{Model}

We consider a homogeneous ensemble of $N$ spin-$1/2$ particles with all-to-all ferromagnetic interactions along $x$ and a uniform magnetic field along $z$ [Fig.~\ref{fig:phenomenon}(a)].
Each spin is subject to local decay and pure dephasing.
We denote the Pauli operators for spin $i$ by $\sigma_i^\alpha$, with $\alpha=x,y,z$.
Setting $\hbar=1$, the Hamiltonian reads
\begin{equation}
	\begin{gathered}
		H=\frac{h_z}{2}\sum_{i=1}^{N}\sigma_i^z-\frac{J}{N}\sum_{i<j}\sigma_i^x\sigma_j^x,\quad J,h_z>0,
	\end{gathered}
	\label{eq:H}
\end{equation}
where $J$ is the ferromagnetic coupling strength and $h_z$ is the magnetic field strength.
The factor $1/N$ ensures that the interaction energy is extensive.
In collective-spin notation, $S_\alpha=\frac12\sum_i\sigma_i^\alpha$, the Hamiltonian is $H=h_zS_z-(2J/N)S_x^2+J/2$.
It is thus a one-axis Lipkin--Meshkov--Glick-type model~\cite{LMG1965,Morrison2008}; the dissipative channels considered here act independently on individual spins.
The combination of this interaction with local spin relaxation also underlies dissipative Ising criticality~\cite{Paz2021}.

The density operator $\rho$ evolves under the Liouvillian superoperator $\mathcal L$ according to
\begin{align}
	\dot\rho
	&=\mathcal L\rho=-i[H,\rho]\nonumber\\
	&\quad+\sum_{i=1}^{N}
	\left(\gamma\mathcal D[\sigma_i^-]
	+\gamma_\phi\mathcal D[\sigma_i^z]\right)\rho,
	\label{eq:lindblad}
\end{align}
where $\gamma>0$ and $\gamma_\phi\geq0$ are the local decay and pure-dephasing coefficients, respectively, and $\sigma_i^-=(\sigma_i^x-i\sigma_i^y)/2$ is the spin-lowering operator.
The Lindblad dissipator associated with a jump operator $L$ is defined as $\mathcal D[L]\rho=L\rho L^\dagger-\{L^\dagger L,\rho\}/2$~\cite{Lindblad1976}.
All Hamiltonian parameters and dissipative coefficients are time independent.
Local decay sets the longitudinal relaxation rate, $\Gamma_1=\gamma$, while both decay and pure dephasing contribute to the transverse relaxation rate, $\Gamma_2=\gamma/2+2\gamma_\phi$.

\subsection{Mean-field dynamics}

We first analyze the dynamics in the thermodynamic limit within the homogeneous mean-field approximation~\cite{Carollo2024}, in which we neglect connected correlations between distinct spins and assume identical local expectation values for all spins.
Defining the Bloch-vector components as $m_\alpha=\avg{M_\alpha}$, where $M_\alpha=\frac{1}{N}\sum_{i=1}^{N}\sigma_i^\alpha$ are the normalized magnetization operators, we obtain the nonlinear Bloch equations
\begin{subequations}\label{eq:bloch}
	\begin{align}
		\dot m_x&=-h_zm_y-\Gamma_2m_x,\\
		\dot m_y&=h_zm_x+2Jm_xm_z-\Gamma_2m_y,\\
		\dot m_z&=-2Jm_xm_y-\Gamma_1(m_z+1).
	\end{align}
\end{subequations}
Figure~\ref{fig:phenomenon} shows how local dephasing turns relaxation into sustained collective motion. Without pure dephasing, the longitudinal magnetization approaches a stationary value [Fig.~\ref{fig:phenomenon}(b)]. Raising $\gamma_\phi$ from zero to $1.5$, with the Hamiltonian, $\Gamma_1$, and initial state unchanged, produces a persistent periodic response [Fig.~\ref{fig:phenomenon}(c)]. The added channel thus sustains oscillations despite enhancing transverse damping. This autonomous motion requires neither coherent driving nor incoherent pumping.

Besides the polarized fixed point $(0,0,-1)$, Eqs.~\eqref{eq:bloch} admit a symmetry-related pair $F_\pm$ when $2h_zJ>h_z^2+\Gamma_2^2$:
\begin{align}
 m_z^F&=-\frac{h_z^2+\Gamma_2^2}{2h_zJ},\nonumber\\
 (m_x^F)^2&=\frac{\Gamma_1(2h_zJ-h_z^2-\Gamma_2^2)}{4\Gamma_2J^2},\nonumber\\
 m_y^F&=-\frac{\Gamma_2}{h_z}m_x^F.
 \label{eq:fixed_points}
\end{align}
Linearizing the original Bloch equations at either member of this pair gives the characteristic equation
\begin{equation}
 \begin{aligned}
 0=(s+2\Gamma_2)\Bigl[&s^2+\Gamma_1s\\
 &+\frac{\Gamma_1}{\Gamma_2}
 (2h_zJ-h_z^2-\Gamma_2^2)\Bigr].
 \end{aligned}
 \label{eq:fixed_stability}
\end{equation}
For $\Gamma_1>0$, all roots have negative real parts throughout the pair's existence domain.
The limit cycle therefore coexists with two locally stable fixed points rather than emerging from their Hopf destabilization.
The asymptotic response depends on which basin of attraction contains the initial state, as in other time-crystal models with non-Hopf transitions or stationary--periodic coexistence~\cite{Cabot2024,Mattes2023}.

\begin{figure}[tbp]
\centering
\includegraphics[width=\columnwidth]{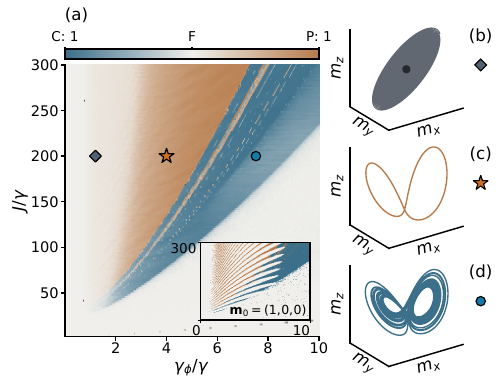}
\caption{Collective responses in the interaction--dephasing plane.
    (a) Response map for $h_z=J$ and fixed local decay $\gamma$, using 200 initial Bloch vectors per grid point.
    Hue indicates the more frequent response category, periodic (P) or chaotic (C), and intensity gives its fraction of all preparations; F marks the stationary center and gray marks a nonzero P/C tie.
    The inset shows the response from the fixed initial Bloch vector $\boldsymbol m(0)=(1,0,0)$ over the same parameter ranges, with the same horizontal and vertical coordinates as the main map.
    Its blue, light-gray, and ochre regions denote C, F, and P responses, respectively.
    (b)--(d) Fixed-point, periodic, and chaotic trajectories from the same sampled initial state at $J/\gamma=200$ and $\gamma_\phi/\gamma=1.2,4.0,7.52$, respectively.
    Diamonds, stars, and circles locate these parameters in (a).
    The three-dimensional axes show per-spin magnetizations, with independently chosen ranges.
    P and C denote periodic and chaotic responses within the observation window, including long-lived transients.
    }
\label{fig:phase}
\end{figure}

To characterize this initial-state dependence, we evolve 200 randomly sampled initial Bloch vectors at each parameter point and construct the dynamical response map in Fig.~\ref{fig:phase}(a), with $h_z=J$ and fixed $\gamma$.
The hue indicates whether periodic or chaotic responses are more frequent, while the color intensity gives the fraction of all 200 trajectories assigned to that category.
Nonstationary responses occupy a finite dephasing window, bounded by predominantly stationary regimes at weaker and stronger dephasing.
Varying the ferromagnetic coupling shifts and reshapes this window, suggesting that sustained motion requires a suitable balance between interactions and dephasing.
Near the boundaries between dynamical regimes, long-lived periodic and chaotic transients also occur.
We include these transients in the corresponding periodic or chaotic response categories according to their behavior within the finite observation window.
Representative fixed-point, periodic, and chaotic trajectories are shown in Figs.~\ref{fig:phase}(b)--\ref{fig:phase}(d), respectively.

A readily prepared initial state is the fully $x$-polarized product state, $\boldsymbol m(0)=(1,0,0)$, obtainable by a global spin rotation of a polarized ensemble~\cite{Bohnet2016}.
The inset in Fig.~\ref{fig:phase}(a) maps the collective response from this fixed initial state, providing a parameter guide for observing periodic oscillations with this preparation.

The Liouvillian $\mathcal L$ is invariant under the $\mathbb Z_2$ parity transformation $\rho\mapsto P\rho P$, where $P=\prod_{i=1}^{N}\sigma_i^z$.
At the mean-field level, this transformation acts as $(m_x,m_y,m_z)\mapsto(-m_x,-m_y,m_z)$.
For the symmetric periodic branch studied here, the parity transformation is equivalent to a half-period time shift along the orbit,
\begin{equation}
	(m_x,m_y,m_z)(t+T/2)=(-m_x,-m_y,m_z)(t),
	\label{eq:orbit_symmetry}
\end{equation}
where $T$ is the full orbital period.
The transverse and longitudinal signals have periods $T$ and $T/2$, corresponding to fundamental angular frequencies $\Omega=2\pi/T$ and $2\Omega$, respectively.
To understand how local dephasing can sustain collective oscillations despite enhancing transverse damping, we next examine the energy balance of the interacting ensemble.

\section{From local dephasing to collective energy feedback}
\label{sec:mechanism}

The physical question is how a channel that damps transverse coherence can sustain collective motion.
We first establish an exact energy balance for the quantum model, then use its homogeneous mean-field limit to prove the necessity of dephasing and identify the delayed feedback that compensates damping.

\subsection{Exact interaction-energy balance}
\label{sec:exact_energy}

Define the transverse pair-correlation operator and a shifted, rescaled energy operator by
\begin{equation}
 \begin{aligned}
 U_N&\equiv\frac{2}{N^2}\sum_{i<j}\sigma_i^x\sigma_j^x
      =M_x^2-\frac IN,\\
 Q_N&\equiv\frac{2H}{N}+h_zI=h_z(M_z+I)-JU_N,
 \end{aligned}
 \label{eq:quantumQ}
\end{equation}
where $I$ is the identity operator.
The local dissipators give
$\mathcal L_{\mathrm{diss}}^\dagger(M_z+I)=-\Gamma_1(M_z+I)$ and
$\mathcal L_{\mathrm{diss}}^\dagger U_N=-2\Gamma_2U_N$.
Since $[H,Q_N]=0$, the coherent contribution to the energy derivative vanishes.
Using $2\Gamma_2=\Gamma_1+4\gamma_\phi$, we obtain the exact identity
\begin{equation}
 \begin{aligned}
  \mathcal L^\dagger Q_N
  &=-\Gamma_1Q_N+4J\gamma_\phi U_N\\
  &=-2\Gamma_2Q_N+4h_z\gamma_\phi(M_z+I).
 \end{aligned}
 \label{eq:quantum_energy_balance}
\end{equation}
The first form separates relaxation of $Q_N$ from an interaction-energy source carried by transverse pair correlations.
The second expresses the same response in terms of longitudinal magnetization and will determine the quantum-mode relations in Sec.~\ref{sec:quantum_energy}.

This identity has a simple microscopic interpretation.
Parallel transverse spins, whether aligned along $+x$ or $-x$, lower the ferromagnetic interaction energy.
Dephasing in the $z$ basis suppresses the local coherence supporting that alignment and reduces the transverse pair correlations.
For $e_N=\avg{H}/N$, its direct contribution is
\begin{equation*}
 \left.\dot e_N\right|_\phi=2J\gamma_\phi\avg{U_N}.
\end{equation*}
Thus, for positive transverse ferromagnetic correlations, dephasing raises the interaction energy even though it does not directly change the local $z$ populations.
The source is state dependent at finite $N$; the exact identity does not assert positive charging for arbitrary pair correlations.
Its role becomes especially transparent in homogeneous mean field, where $\avg{U_N}\to m_x^2$.

\subsection{Mean-field energy balance and the necessity of dephasing}

The Hamiltonian energy per spin provides the natural starting point.
Under homogeneous factorization, $\avg{U_N}\to m_x^2$ and $\avg{Q_N}\to Q$.
The Hamiltonian in Eq.~\eqref{eq:H} gives
\begin{equation*}
e\equiv\lim_{N\to\infty}\frac{\avg{H}}{N}
=\frac{h_z}{2}m_z-\frac{J}{2}m_x^2.
\end{equation*}
The factor $1/2$ in the interaction term follows from the $N(N-1)/2$ distinct pairs together with the $1/N$ Kac normalization.
To remove the constant offset in the local-decay contribution to the energy balance, we introduce the shifted and rescaled energy variable
\begin{equation}
	\begin{aligned}
	Q&\equiv2\left(e+\frac{h_z}{2}\right)=h_z(m_z+1)-Jm_x^2.
	\end{aligned}
	\label{eq:Q}
\end{equation}
With this choice, local decay contributes the simple relaxation term $-\Gamma_1Q$.
Substituting Eqs.~\eqref{eq:bloch} into the time derivative of Eq.~\eqref{eq:Q} gives
\begin{equation}
 \begin{aligned}
 \dot Q&=h_z\dot m_z-2Jm_x\dot m_x\\
 &=-\Gamma_1h_z(m_z+1)+2J\Gamma_2m_x^2\\
 &=-\Gamma_1Q+4J\gamma_\phi m_x^2.
 \end{aligned}
 \label{eq:Qdot}
\end{equation}
This is the mean-field limit of the exact identity~\eqref{eq:quantum_energy_balance}.
Local decay relaxes $Q$ toward zero, whereas dephasing supplies a positive source proportional to $m_x^2$. On the periodic response considered below, $Q\geq0$, so Eq.~\eqref{eq:Qdot} directly separates interaction-energy charging from energy loss.

The same variable feeds back on the transverse motion.
Using
\begin{equation*}
 \begin{aligned}
 m_y&=-\frac{\dot m_x+\Gamma_2m_x}{h_z},\\
 m_z&=\frac{Q+Jm_x^2}{h_z}-1
 \end{aligned}
\end{equation*}
to eliminate $m_y$ and $m_z$ from the Bloch equations gives
\begin{equation}
 \begin{aligned}
 \ddot m_x+2\Gamma_2\dot m_x
 &=-\left(h_z^2+\Gamma_2^2-2h_zJ\right)m_x\\
 &\quad-2J^2m_x^3-2JQm_x.
 \end{aligned}
 \label{eq:oscillator}
\end{equation}
Equation~\eqref{eq:oscillator} describes a damped nonlinear oscillator coupled to the relaxing energy variable $Q$.
Thus $m_x^2$ drives the energy variable through Eq.~\eqref{eq:Qdot}, while the energy variable acts back through the force $-2JQm_x$ in Eq.~\eqref{eq:oscillator}.

This form provides a direct proof that pure dephasing is necessary for periodic motion in the homogeneous mean-field dynamics of this model when $\Gamma_1>0$.
At $\gamma_\phi=0$, Eq.~\eqref{eq:Qdot} gives $Q(t)=Q(0)e^{-\Gamma_1t}$, so periodicity requires $Q\equiv0$.
When Eq.~\eqref{eq:oscillator} is multiplied by $\dot m_x$ and integrated over a period $T$, all conservative terms reduce to endpoint differences and vanish.
Hence
\begin{equation}
 2\Gamma_2\int_0^T\dot m_x^2\,\dd t=0.
 \label{eq:no_dephasing_cycle}
\end{equation}
Because $\Gamma_2=\Gamma_1/2>0$, $m_x$ must be constant; the relations above then imply that $m_y$ and $m_z$ are constant as well.
With nonzero local decay, pure dephasing is therefore necessary for any periodic orbit of the homogeneous mean-field dynamics.

\subsection{Delayed feedback and sustained oscillations}

\begin{figure}[tbp]
\centering
\includegraphics[width=\columnwidth]{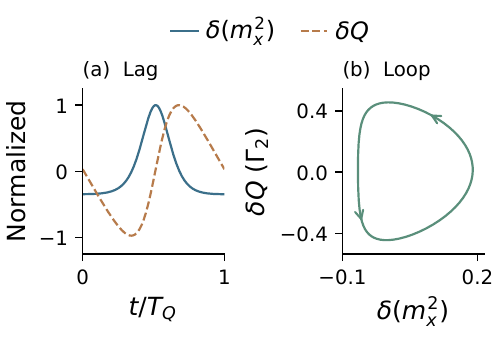}
\caption{Delayed mean-field energy feedback.
    (a) Centered $m_x^2$ and $Q$, each divided by its maximum absolute deviation, over one feedback period $T_Q=T/2$.
    The energy response lags the transverse pair order.
    (b) The same orbit in centered $(m_x^2,Q)$ coordinates; arrows indicate increasing time.
    Because the lag makes $Q$ different on the increasing and decreasing branches of $m_x^2$, the trajectory encloses an oriented loop.
    The loop gives the cycle-integrated feedback contribution in Eq.~\eqref{eq:feedback_loop}.
    Parameters are $J/\Gamma_2\approx31.6$, $h_z/\Gamma_2\approx31.6$, $\Gamma_1/\Gamma_2=0.3$, and $\gamma_\phi/\Gamma_2=0.425$, with $\Omega/\Gamma_2\approx6.6$.}
\label{fig:mechanism}
\end{figure}

Charging and loss can balance at a fixed point, so their presence alone does not explain sustained motion.
What distinguishes the periodic response is that the energy follows the transverse order with a finite relaxation time and therefore acts back with a delay.
Solving Eq.~\eqref{eq:Qdot} with an integrating factor from an arbitrary initial time $t_0$ gives
\begin{equation}
 \begin{aligned}
 Q(t)&=e^{-\Gamma_1(t-t_0)}Q(t_0)\\
 &\quad+4J\gamma_\phi\int_{t_0}^{t}
 e^{-\Gamma_1(t-t')}m_x^2(t')\,\dd t'.
 \end{aligned}
 \label{eq:Qinitial}
\end{equation}
For the bounded long-time periodic response, the initial term has decayed.
Setting $s=t-t'$ gives
\begin{equation}
 Q(t)=4J\gamma_\phi\int_0^\infty
 e^{-\Gamma_1s}m_x^2(t-s)\,\dd s.
 \label{eq:Qmemory}
\end{equation}
The kernel weights the recent history of $m_x^2$ over the memory time $\Gamma_1^{-1}$.
It also makes $Q\geq0$ on the periodic orbit.
Averaging Eq.~\eqref{eq:Qmemory} over a period $T_Q$ shared by $Q$ and $m_x^2$ gives the global charging--loss balance
\begin{equation}
 \begin{aligned}
 \Gamma_1\avg{Q}_{T_Q}&=4J\gamma_\phi\avg{m_x^2}_{T_Q},\\
 \avg{f}_{T_Q}&\equiv\frac1{T_Q}\int_0^{T_Q}f(t)\,\dd t.
 \end{aligned}
 \label{eq:charging_balance}
\end{equation}

The half-period symmetry in Eq.~\eqref{eq:orbit_symmetry} makes $Q$ and $m_x^2$ repeat after $T_Q=T/2$.

This balance fixes the cycle-averaged energy, while the relative timing of $Q$ and $m_x^2$ within the cycle is determined by the memory kernel in Eq.~\eqref{eq:Qmemory}.
To quantify this temporal response, we expand both quantities in Fourier harmonics and evaluate the memory integral for each component with time dependence $e^{\ii\nu t}$, obtaining
\begin{equation}
 \begin{aligned}
 Q_\nu&=\frac{4J\gamma_\phi}{\Gamma_1+\ii\nu}(m_x^2)_\nu\\
 &=\frac{4J\gamma_\phi}{\sqrt{\Gamma_1^2+\nu^2}}
 e^{-\ii\delta_\nu}(m_x^2)_\nu,\\
 \delta_\nu&=\arctan\!\left(\frac{\nu}{\Gamma_1}\right).
 \end{aligned}
 \label{eq:energy_response}
\end{equation}
At $\nu=0$, this relation reproduces the cycle-averaged balance in Eq.~\eqref{eq:charging_balance}.

At fixed $J$, $\Gamma_1$, and $\nu>0$, the response gain is proportional to $\gamma_\phi$, whereas the finite energy-relaxation time produces the phase lag $\delta_\nu$.

Figure~\ref{fig:mechanism}(a) resolves this phase lag in the time domain, while Fig.~\ref{fig:mechanism}(b) displays the associated loop in $(m_x^2,Q)$.
At a fixed value of $m_x^2$, $Q$ is smaller on one branch of the cycle and larger on the other.
Since the transverse feedback force is $-2JQm_x$, this asymmetry produces a nonzero net contribution over a closed cycle.
\begin{figure*}[t]
	\centering
	\includegraphics[width=\textwidth]{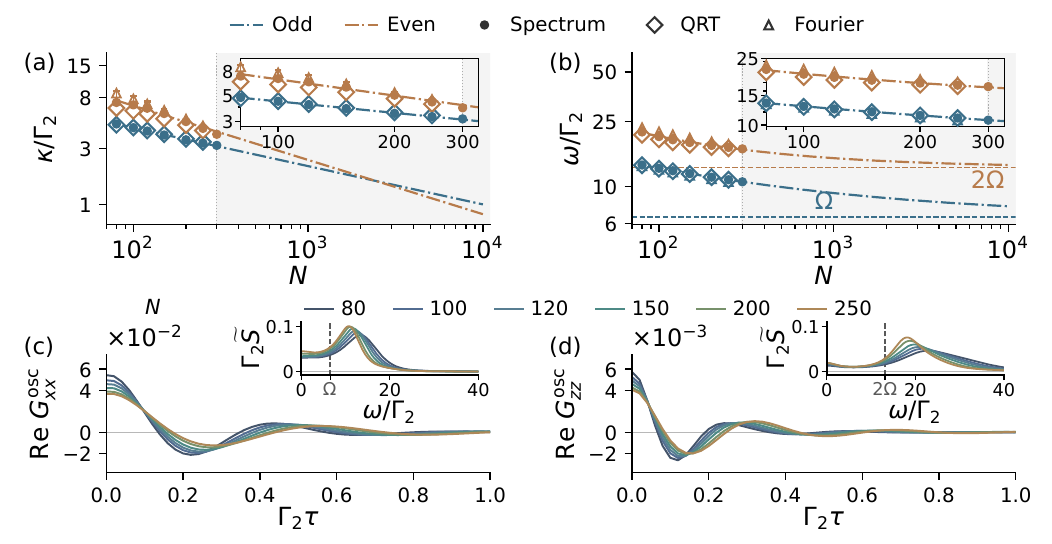}
	\caption{Finite-size scaling of collective quantum oscillations.
	    (a,b) Decay rates and frequencies of odd- and even-parity modes.
	    Circles denote Liouvillian eigenvalues, diamonds QRT estimates, and triangles Fourier half-widths in (a) and peak positions in (b). Error bars show the range of estimates obtained using different fitting windows.
	    Dash-dotted curves are fits with fixed large-$N$ limits $\kappa=0$ and $\omega=\Omega,2\Omega$; dashed lines mark the mean-field frequencies.
	    Shading indicates extrapolation beyond $N=300$, and insets enlarge the small-$N$ region.
	    (c,d) Real parts of the transverse and longitudinal oscillatory correlations, retaining their absolute per-spin amplitudes.
	    Insets show symmetric spectra normalized to unit area over the nonnegative frequency grid; colors distinguish system sizes.
	    Parameters are those of Fig.~\ref{fig:mechanism}.}
	\label{fig:scaling}
	\label{fig:correlations}
\end{figure*}

Multiplying Eq.~\eqref{eq:oscillator} by $\dot m_x$ and integrating over one feedback period of the symmetric orbit gives the global balance
\begin{equation}
 2\Gamma_2\int_0^{T_Q}\dot m_x^2\,\dd t
 =-J\oint Q\,\dd(m_x^2)>0.
 \label{eq:feedback_loop}
\end{equation}
The positive sign follows directly by setting $u=m_x^2$ and using Eq.~\eqref{eq:Qdot}. Integration by parts over a cycle gives
\begin{align}
 -J\avg{Q\dot u}_{T_Q}
 &=J\avg{\dot Q u}_{T_Q}\nonumber\\
 &=\frac{\avg{\dot Q^2}_{T_Q}}{4\gamma_\phi}>0,
 \label{eq:positive_feedback}
\end{align}
where $\avg{Q\dot Q}_{T_Q}=0$ by periodicity. Thus the delayed response supplies strictly positive feedback work on a periodic orbit.
The left-hand side is the transverse damping accumulated over the cycle, while the right-hand side is the generalized work of the internal feedback force.
Equation~\eqref{eq:feedback_loop} is therefore a cycle-integrated description of sustained motion: the oriented loop in Fig.~\ref{fig:mechanism}(b) exactly compensates the damping of the reduced transverse dynamics.
The physical energy charging and loss themselves remain those in Eq.~\eqref{eq:Qdot}.

If the energy followed $m_x^2$ instantaneously, $Q=(4J\gamma_\phi/\Gamma_1)m_x^2$, the path in the $(m_x^2,Q)$ plane would retrace a line and the contour integral in Eq.~\eqref{eq:feedback_loop} would vanish.
Dephasing-induced charging is therefore necessary, but finite response time is what converts that charging into the delayed feedback capable of sustaining the cycle.

This mechanism characterizes the periodic orbit once it exists; its emergence, coexistence with stationary states, and basin structure are governed by the full nonlinear dynamics of Eqs.~\eqref{eq:bloch}.

Together, the necessity proof and the cycle balance explain how local dephasing, acting through ferromagnetic interactions, can sustain collective oscillations despite enhancing transverse damping.
The exact identity~\eqref{eq:quantum_energy_balance} carries this energy-response structure into the quantum analysis below.

This interaction-energy mechanism complements symmetry-based routes to nonstationarity and synchronization~\cite{Buca2019,BucaJaksch2019,Tindall2020,BucaAlgebra2022,AlaeianBuca2022} and dephasing-assisted temporal order under periodic driving~\cite{Sakurai2021}.

\section{Finite-size quantum dynamics and time-crystalline order}
\label{sec:quantum}
Continuous time-crystalline order requires nondecaying collective oscillations at a nonzero frequency in the thermodynamic limit, reflecting spontaneous breaking of continuous time-translation symmetry.
To characterize this order beyond mean field, we identify the oscillatory modes of the Liouvillian $\mathcal L$ that contribute to collective two-time correlations and analyze three aspects of their system-size dependence: decay rates, oscillation frequencies, and collective oscillatory weights.
We then investigate the energy response associated with these modes, connecting the finite-size quantum dynamics to the feedback mechanism developed in Sec.~\ref{sec:mechanism}.

\subsection{Finite-size scaling and temporal order}
\label{sec:quantum_scaling}

Exploiting permutational invariance, we construct a Dicke-block representation of the Liouvillian $\mathcal L$ defined in Eq.~\eqref{eq:lindblad} and analyze its eigenmodes~\cite{Shammah2018}.
A nonstationary right eigenmatrix $R_k$ satisfies
\begin{equation}
	\mathcal L R_k=\lambda_kR_k,\qquad
	\lambda_k=-\kappa_k+\ii\omega_k\neq0,
	\label{eq:mode}
\end{equation}
where $\kappa_k$ and $\omega_k$ denote the decay rate and angular frequency, respectively.
Trace preservation implies $\Tr R_k=0$ for every nonzero eigenvalue.
An oscillatory mode ($\omega_k\neq0$) evolves with time dependence $e^{-\kappa_k t+\ii\omega_k t}$ and has a lifetime $\kappa_k^{-1}$ for $\kappa_k>0$.

The parity symmetry introduced in Sec.~\ref{sec:model} allows the modes to be classified into odd and even sectors.
The symmetric mean-field orbit in Eq.~\eqref{eq:orbit_symmetry} provides the reference frequencies $\Omega$ and $2\Omega$ for the transverse and longitudinal oscillations, respectively.
To assess continuous time-crystalline order, we examine how the corresponding oscillatory branches evolve with system size.
Over the studied range $N=80$--$300$, both branches become longer lived, with the power-law fits in Fig.~\ref{fig:scaling}(a) yielding $\kappa_{\mathrm{odd}}\propto N^{-0.323}$ and $\kappa_{\mathrm{even}}\propto N^{-0.458}$.
Their frequencies move toward the mean-field values $\Omega$ and $2\Omega$, respectively, as shown in Fig.~\ref{fig:scaling}(b).

To establish how these modes contribute to collective fluctuations, we calculate stationary connected two-time correlations using the quantum regression theorem (QRT)~\cite{Lax1963,GardinerZoller2004}.
For observables $A$ and $B$ and delay $\tau\geq0$, we define
\begin{equation}
	\begin{aligned}
	G_{AB}(\tau)
	&=\Tr\!\left[\delta A(\tau)\delta B\rho_\stead\right]\\
	&=\Tr\!\left[\delta A\,e^{\mathcal L\tau}
	 (\delta B\rho_\stead)\right],
	\end{aligned}
	\label{eq:QRT}
\end{equation}
where $\rho_\stead$ is the stationary density operator, $I$ is the identity operator, $\delta A=A-\avg{A}_\stead I$, and $\avg{A}_\stead=\Tr(A\rho_\stead)$; $\delta B$ is defined analogously.
For $A=B=M_\alpha$, we write $G_{\alpha\alpha}$, with the per-spin normalization already included in $M_\alpha$.
These correlations probe collective fluctuations even when the stationary mean magnetization is time independent.

For a diagonalizable Liouvillian with a complete eigenbasis, the magnetization autocorrelations can be expressed as a sum of Liouvillian modes,
\begin{equation}
	G_{\alpha\alpha}(\tau)
	=\sum_{k\ne0}w_{\alpha k}e^{\lambda_k\tau},
	\label{eq:spectralexpansion}
\end{equation}
where $w_{\alpha k}$ denotes the generally complex spectral weight of mode $k$ in the $\alpha$ channel.
For a parity-invariant stationary state, only odd-parity modes contribute to $G_{xx}$, whereas only even-parity modes contribute to $G_{zz}$.

Slow Liouvillian modes describe relaxation and metastable dynamics~\cite{Macieszczak2016,Rose2016,Macieszczak2021,Minganti2018}. To isolate the collective oscillations, we separate each correlation into a slowly decaying background and an oscillatory contribution,
\begin{equation}
	G_{\alpha\alpha}(\tau)
	=B_\alpha(\tau)+G^{\osc}_{\alpha\alpha}(\tau).
	\label{eq:backgroundsplit}
\end{equation}
The backgrounds $B_x(\tau)$ and $B_z(\tau)$ are obtained from fits to the long-delay correlations. Their assignments to nonoscillatory transverse and low-frequency longitudinal modes, respectively, are supported by independently computed Liouvillian spectra.
After background subtraction, the transverse and longitudinal correlations in Fig.~\ref{fig:scaling}(c,d) exhibit damped oscillations whose decay rates and frequencies track the odd- and even-parity branches identified above.
This correspondence connects the spectral modes to oscillations of the collective magnetization.
As $N$ increases, these correlations persist over longer delays, while the spectral peaks shown in the insets narrow and shift toward the mean-field frequencies.

\begin{figure}[t]
\centering
\includegraphics[width=\columnwidth]{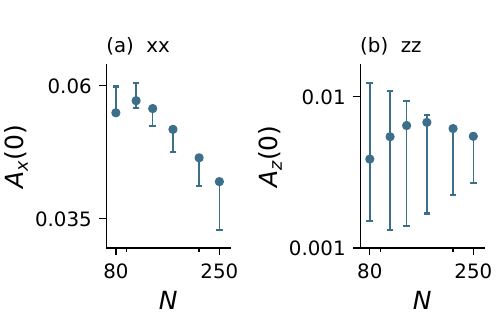}
\caption{Absolute oscillatory amplitudes.
    (a) Transverse and (b) longitudinal envelope amplitudes $A_\alpha(0)$ in Eq.~\eqref{eq:realfit}.
    Circles show the fitted amplitudes; error bars span the estimates obtained using different fitting windows.
    The amplitudes retain their absolute per-spin scale; the error bars quantify fitting-window sensitivity.
    Parameters are those of Fig.~\ref{fig:mechanism}.}
\label{fig:amplitudes}
\end{figure}

To quantify the collective oscillatory weights on the absolute per-spin scale, we fit the extracted correlations to
\begin{equation}
 \begin{aligned}
 \operatorname{Re}G^{\osc}_{\alpha\alpha}(\tau)
 &\simeq A_\alpha(0)e^{-\kappa_\alpha\tau}\\
 &\quad\times\cos(\omega_\alpha\tau+\varphi_\alpha),
 \qquad \alpha=x,z.
 \end{aligned}
 \label{eq:realfit}
\end{equation}
Here $A_\alpha(0)$ is the fitted oscillatory envelope amplitude extrapolated to zero delay; $\kappa_\alpha$, $\omega_\alpha$, and $\varphi_\alpha$ are the decay rate, angular frequency, and phase offset of the oscillatory contribution.
Figure~\ref{fig:amplitudes} shows these amplitudes for $N=80$--$250$.
The transverse central estimate changes from approximately $5.4\times10^{-2}$ to $4.1\times10^{-2}$.
The longitudinal estimate changes from $3.9\times10^{-3}$ to $5.5\times10^{-3}$, reaching $6.8\times10^{-3}$ at $N=150$.
Within each channel, the oscillatory weight remains of comparable absolute magnitude across the studied sizes.
The error bars span the amplitude estimates obtained using different fitting windows, with the larger longitudinal spans reflecting sensitivity to the background separation and oscillatory fitting interval.

Continuous time-crystalline order requires an oscillatory contribution whose decay rate vanishes in the thermodynamic limit, while its frequency and weight in the per-spin collective correlation remain nonzero~\cite{Iemini2018,Booker2020,Cabot2024}.
Here the thermodynamic limit is taken before the long-delay limit.
The algebraic decrease of the fitted decay rates, convergence toward nonzero mean-field frequencies, and comparable per-spin oscillatory amplitudes jointly support dephasing-induced continuous time-crystalline order.

\subsection{Energy response of collective oscillations}
\label{sec:quantum_energy}

The exact identity~\eqref{eq:quantum_energy_balance} now connects the collective oscillatory modes to the interaction-energy feedback of Sec.~\ref{sec:mechanism}.
Because the identity holds at finite $N$, this connection does not require factorization of the pair correlations.

To resolve this energy response within the oscillatory modes identified in Sec.~\ref{sec:quantum_scaling}, we define
\[
 q_k=\Tr(Q_NR_k),\quad
 u_k=\Tr(U_NR_k),\quad
 z_k=\Tr(M_zR_k).
\]
Since $Q_N$, $U_N$, and $M_z$ are parity even, we evaluate these projections for the even-parity oscillatory branch.
For nonzero $u_k$ and $z_k$, projecting Eq.~\eqref{eq:quantum_energy_balance} onto $R_k$ and using $\Tr R_k=0$ gives the relative energy responses
\begin{equation}
	\begin{aligned}
		\frac{q_k}{u_k}
		&=\frac{4J\gamma_\phi}{\Gamma_1+\lambda_k},\\
		\frac{q_k}{h_z z_k}
		&=\frac{4\gamma_\phi}{2\Gamma_2+\lambda_k}.
	\end{aligned}
	\label{eq:mode_response}
\end{equation}
These ratios are independent of the normalization of $R_k$.
The first extends the mean-field harmonic response in Eq.~\eqref{eq:energy_response} to a quantum oscillatory mode, with $\ii\nu$ replaced by $\lambda_k=-\kappa_k+\ii\omega_k$.
The finite-size decay rate therefore enters the response denominator together with the oscillation frequency, determining both the relative energy amplitude and its phase.
The second ratio expresses the same energy response relative to the longitudinal magnetization.

\begin{figure}[t]
\centering
\includegraphics[width=\columnwidth]{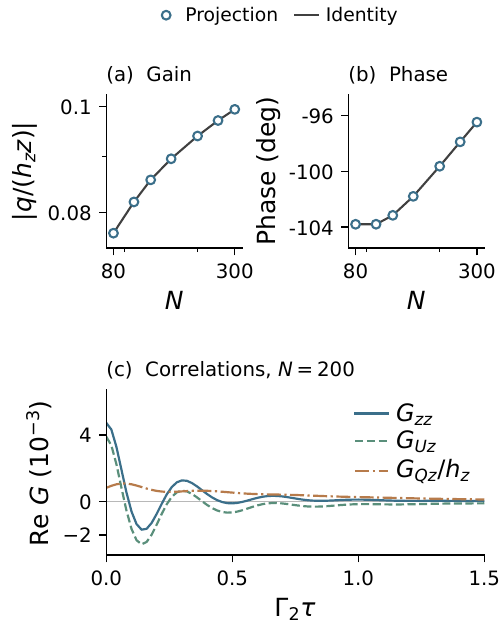}
\caption{Quantum energy response.
    (a,b) Gain and phase of $q_k/(h_zz_k)$ for even oscillatory modes.
    Circles are direct eigenmatrix projections; the line connects the values obtained from Eq.~\eqref{eq:mode_response} at the same eigenvalues.
    (c) Real parts of the longitudinal, pair--magnetization, and energy--magnetization correlations at $N=200$.
    The latter two are reconstructed from the full complex $G_{zz}$ and Eq.~\eqref{eq:energy_correlation_response}, with stationary initial covariances.
    Parameters are those of Fig.~\ref{fig:mechanism}.}
\label{fig:quantumresponse}
\end{figure}

Figure~\ref{fig:quantumresponse}(a,b) shows the relative energy amplitude and phase obtained from the even-parity eigenmodes, in agreement with Eq.~\eqref{eq:mode_response}.
Across the studied sizes, the relative energy amplitude increases, accompanied by an overall reduction in the magnitude of the phase lag.
For $\omega_k>0$ and $\gamma_\phi>0$, the energy phase relative to the longitudinal magnetization is
\begin{equation}
	\arg\!\left(\frac{q_k}{h_z z_k}\right)
	=-\arg\!\left(2\Gamma_2-\kappa_k+\ii\omega_k\right).
	\label{eq:mode_phase}
\end{equation}
The phases below $-90^\circ$ in Fig.~\ref{fig:quantumresponse}(b) arise from $\kappa_k>2\Gamma_2$, which makes the real part of the response denominator negative.
The phase shift thus reflects the energy response within a decaying collective oscillation.
As $\kappa_k$ approaches zero, Eq.~\eqref{eq:mode_response} reduces to the purely oscillatory response, connecting the finite-size modes to the harmonic description in Sec.~\ref{sec:mechanism}.

The same energy-response relation also governs the stationary two-time correlations.
Using Eq.~\eqref{eq:QRT}, we denote the energy--magnetization and pair--magnetization correlations by $G_{Qz}$ and $G_{Uz}$, corresponding to $(A,B)=(Q_N,M_z)$ and $(U_N,M_z)$, respectively.
Equation~\eqref{eq:quantum_energy_balance} gives
\begin{equation}
	\frac{\mathrm dG_{Qz}}{\mathrm d\tau}
	=-2\Gamma_2G_{Qz}+4h_z\gamma_\phi G_{zz}.
	\label{eq:energy_correlation}
\end{equation}
Its solution separates the relaxation of the initial covariance from the response to longitudinal correlations,
\begin{equation}
	\begin{aligned}
		G_{Qz}(\tau)
		&=e^{-2\Gamma_2\tau}G_{Qz}(0)\\
		&\quad+4h_z\gamma_\phi\int_0^\tau
		e^{-2\Gamma_2(\tau-s)}G_{zz}(s)\,\mathrm ds.
	\end{aligned}
	\label{eq:energy_correlation_response}
\end{equation}
The convolution makes explicit how the energy correlation responds to the history of the longitudinal collective fluctuations.
The pair--magnetization correlation follows from $G_{Uz}=(h_zG_{zz}-G_{Qz})/J$.
Figure~\ref{fig:quantumresponse}(c) shows these energy and pair responses at $N=200$, reconstructed from the full longitudinal correlation and the stationary initial covariance.
Together with the mode projections, these time-domain responses show how the energy-response structure developed in Sec.~\ref{sec:mechanism} is encoded in the collective quantum oscillations.

\section{Intertwined time-crystalline dynamics and metastability}
\label{sec:metastability}

The mean-field limit cycle coexists with two stable fixed points. We now examine how intrinsic fluctuations connect these attractors, first through semiclassical trajectories and then through fixed-point occupation statistics. Figure~\ref{fig:semiclassical_switching} resolves repeated capture near fixed points and reactivation of collective oscillations within this three-region dynamics.

\subsection{Noise-induced switching between stationary and oscillatory states}
\label{sec:semiclassical_switching}

\begin{figure*}[tp]
	\centering
	\includegraphics[width=\textwidth]{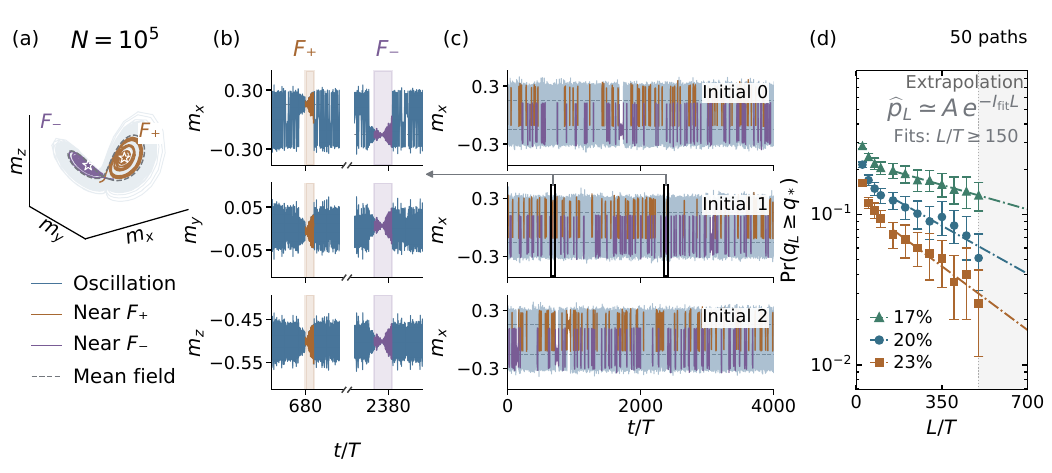}
	\caption{Intertwined oscillatory and stationary metastability at $N=10^5$.
	    (a) Magnetization trajectory over $650\leq t/T\leq2410$; the gray dashed curve and stars denote the mean-field limit cycle and stable fixed points.
	    (b) Enlarged magnetization traces showing fluctuation-induced fixed-point capture and recovery of collective oscillations.
	    (c) Complete $4000T$ traces of $m_x(t)$ for three initial conditions; black boxes in Initial~1 mark the windows shown in (b).
	    Orange and purple identify visits near $F_+$ and $F_-$.
	    (d) Probabilities that the fixed-point occupation fraction exceeds $17\%$, $20\%$, and $23\%$, estimated from $50$ trajectories after discarding the first $128T$.
	    Error bars indicate $95\%$ whole-trajectory bootstrap intervals.
	    Dash-dotted lines show fits of the form $\widehat p_L\simeq A\exp(-I_{\mathrm{fit}}L)$ over $150\leq L/T\leq500$.
	    In increasing threshold order, $(A,I_{\mathrm{fit}}T)\simeq(0.228,0.00107)$, $(0.175,0.00209)$, and $(0.120,0.00278)$.
	    Gray shading marks extrapolation over $500$--$700T$.
	    Parameters: $J=h_z=250$, $\Gamma_1=1$, and $\gamma_\phi=5$; the initial magnetization in (a,b) is approximately $(0.09,-0.26,0.15)$.
	    $T$ is the full mean-field period.}
	\label{fig:semiclassical_switching}
\end{figure*}

The mean-field periodic attractor identified in Sec.~\ref{sec:model} coexists with two symmetry-related stable fixed points, $F_+$ and $F_-$.
The labels follow the sign of $m_x^F$ in Eq.~\eqref{eq:fixed_points}: $m_x^F>0$ at $F_+$ and $m_x^F<0$ at $F_-$.
To resolve how fluctuations connect these coexisting attractors, we describe the magnetization using a semiclassical diffusion approximation that combines the mean-field drift with Gaussian noise from the local dissipative channels~\cite{GelhausenBuchhold2018,vanKampen2007}:
\begin{equation}
	\mathrm d\boldsymbol m
	=\boldsymbol F(\boldsymbol m)\,\mathrm dt
	+\frac{1}{\sqrt N}\boldsymbol B(\boldsymbol m)\,
	\mathrm d\boldsymbol W_t.
	\label{eq:stochastic_magnetization}
\end{equation}
Equation~\eqref{eq:stochastic_magnetization} is interpreted in the It\^o sense.
Here $\boldsymbol m=(m_x,m_y,m_z)$ denotes the random magnetization in a single semiclassical realization, $\boldsymbol F$ is the vector field in Eqs.~\eqref{eq:bloch}, and $\boldsymbol W_t$ contains three independent Wiener processes satisfying $\mathbb E[\mathrm dW_\alpha\,\mathrm dW_\beta] =\delta_{\alpha\beta}\,\mathrm dt$.
The expectation $\mathbb E[\cdot]$ averages over stochastic realizations.
The factor $N^{-1/2}$ reflects the $1/N$ scaling of the noise variance with system size, resulting from independent local noise contributions to the normalized collective magnetization.
This collective diffusion model resolves fluctuation-driven excursions around the mean-field dynamics; it is not an exact quantum-jump unraveling of Eq.~\eqref{eq:lindblad}.
Related phase-space approaches provide complementary tools for large dissipative spin ensembles~\cite{Huber2022,Mink2022}.

We obtain the diffusion matrix by evaluating the local dissipators in the symmetrized second-moment equations and subtracting the drift contributions~\cite{GelhausenBuchhold2018,Xiang2026}:
\begin{equation}
	\begin{aligned}
		\boldsymbol B\boldsymbol B^{\mathsf T}
		&=\boldsymbol C_0(\boldsymbol m),\\
		\boldsymbol C_0(\boldsymbol m)
		&=
		\begin{pmatrix}
			2\Gamma_2 & 0 & \Gamma_1m_x\\
			0 & 2\Gamma_2 & \Gamma_1m_y\\
			\Gamma_1m_x & \Gamma_1m_y & 2\Gamma_1(1+m_z)
		\end{pmatrix}.
	\end{aligned}
	\label{eq:semiclassical_diffusion}
\end{equation}

For the parameters in Fig.~\ref{fig:semiclassical_switching}, we initialize the dynamics in the mean-field limit-cycle basin and follow trajectories at $N=10^5$. Here $T$ and $\Omega=2\pi/T$ refer to the noiseless cycle. Figures~\ref{fig:semiclassical_switching}(a,b) show repeated alternation between oscillatory motion and residence near either fixed point. Capture is temporary: intrinsic fluctuations can reactivate oscillations, as the complete $4000T$ records for three initial conditions confirm [Fig.~\ref{fig:semiclassical_switching}(c)]. We also observe reactivation from an initial state placed directly at a stable mean-field fixed point.

The limit-cycle region has a dual role. It supports many rapid oscillations as a metastable dynamical region, and it provides an observed intermediate channel between the stationary regions: escape from one fixed-point neighborhood can be followed by an extended oscillatory episode and capture near the other. These paths intertwine fast temporal order with slower transitions between regions. This is distinct from phase diffusion, which can erase phase memory without leaving the oscillatory region; reactivation likewise need not restore the earlier phase. The observed switching channel involves the cycle's neighborhood, without implying that every noisy path must cross the deterministic orbit.

Coexistence and switching are established phenomena, but the relevant precedents involve different dynamical structures. Stationary-state switching has been studied in Rydberg, photonic, and Dicke systems~\cite{LeeJumps2012,Letscher2017,Casteels2017,GelhausenBuchhold2018,Xiang2026}. In particular, Stitely \textit{et al.} have resolved switching among a normal state and two symmetry-related superradiant states, with oscillatory dynamics studied in a separate regime~\cite{Stitely2020}. Mattes \textit{et al.} have established stationary--limit-cycle coexistence and hysteresis~\cite{Mattes2023}; quantum-trajectory intermittency at a time-crystal transition~\cite{Cabot2023} and phase-diffusive versus activated relaxation in distinct oscillatory and stationary regimes~\cite{Cabot2024} have also been characterized. Here the distinguishing structure is the recurrent participation of a \emph{dephasing-sustained limit cycle} in switching between two stationary regions at fixed parameters. Temporal order is therefore part of an intermediate metastable episode, not merely a separate phase or a transient preceding stationary relaxation.

\subsection{Temporal statistics of metastable intermittency}
\label{sec:occupation_statistics}

To quantify this intermittency, we define the fixed-point occupation fraction as the fraction of an observation window spent in identified residence intervals near either $F_+$ or $F_-$:
\begin{equation}
	q_L(t_0)=\frac{1}{L}\int_{t_0}^{t_0+L}\chi_F(t)\,\mathrm dt,
	\label{eq:fixed_point_occupation}
\end{equation}
where $\chi_F$ equals one during identified fixed-point residence intervals and zero otherwise.
The remaining intervals predominantly correspond to limit-cycle-associated oscillatory motion, including noise-driven excursions away from the deterministic cycle; they also contain a small number of unresolved brief fixed-point captures.

At fixed $N=10^5$, Fig.~\ref{fig:semiclassical_switching}(d) examines how the probability of exceeding a prescribed occupation threshold decreases with observation time.
The thresholds of $17\%$, $20\%$, and $23\%$ all exceed the sample mean fixed-point occupation fraction of approximately $12.8\%$.
Over the fitting range $150\leq L/T\leq500$, these upper-tail probabilities exhibit an approximately exponential decrease, consistent with a temporal large-deviation description~\cite{Garrahan2010,Lesanovsky2013,Touchette2018,Angeletti2016}.
The effective decay rate increases with the threshold.
Thus, sustaining an unusually large fixed-point occupation fraction over a longer observation time becomes increasingly unlikely.
This is a statement about occupation-time fluctuations at fixed $N$, rather than a measurement of escape-rate scaling with system size.

The occupation statistic complements the trajectory geometry.
The former quantifies how much of a record is interrupted by fixed-point residence, whereas the latter shows how oscillations resume and participate in switching.
Together, they identify a recurrent oscillatory metastable region embedded in broader switching dynamics.
Resolving residence times and transition paths could connect this phase-and-region description to effective metastable dynamics and its trajectory dependence~\cite{Macieszczak2021,Brown2024}. The long oscillatory intervals at $N=10^5$ provide windows for continuous observation, which we examine next.

\section{Observability and robustness}
\label{sec:observability}

The preceding sections distinguish the mechanism sustaining temporal order, its quantum spectral signatures, and its fluctuating realization among coexisting attractors.
We now ask how this dynamics appears in an experimental record.
Using the semiclassical trajectories of Sec.~\ref{sec:metastability}, we examine photon-counting readout of the longitudinal oscillations, then test the response to fluctuations of the interaction and local dissipative rates.

\subsection{Continuous photon readout}
\label{sec:photon_readout}

\begin{figure}[tbp]
\centering
\includegraphics[width=\columnwidth]{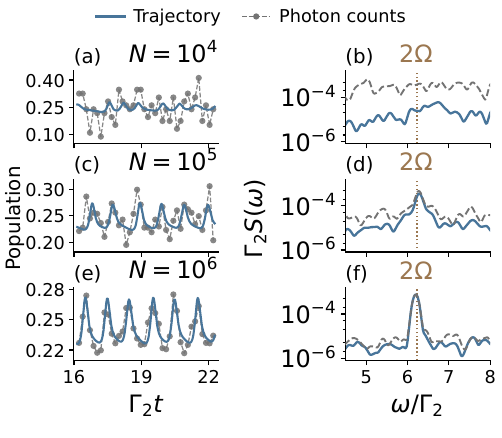}
\caption{Single-record photon readout at $N=10^4$, $10^5$, and $10^6$ (rows).
    (a,c,e) Population $p_\uparrow(t)$ (blue) and count-inferred bin-averaged populations $\hat p_{\uparrow,b}$ (gray markers and dashed lines).
    (b,d,f) Corresponding spectra $S_{\mathrm{traj}}$ (blue) and $S_{\mathrm{count}}$ (gray dashed); dotted lines mark $2\Omega$.
    The efficiency is $\eta=0.3$ and the bin width $\Delta t=0.08T\simeq0.16T_2$, where $T_2=1/\Gamma_2$.
    Parameters and initial magnetization are those of Fig.~\ref{fig:semiclassical_switching}.}
\label{fig:photonreadout}
\end{figure}

We examine how clearly the longitudinal collective oscillations can be resolved in continuous photon-counting records.
Continuous optical monitoring has revealed time-crystalline motion in atom--cavity experiments~\cite{Kongkhambut2022,Cosme2025}.
In the present model, radiative local spin decay provides a direct probe of the excited-state population fraction, $p_\uparrow(t)=[1+m_z(t)]/2$.
For photons from independent local decay channels counted with a common efficiency $\eta$, the mean detection rate along a given semiclassical trajectory is~\cite{Landi2024}
\begin{equation}
	R_{\mathrm{det}}(t)
	=\eta\Gamma_1N p_\uparrow(t).
	\label{eq:photon_rate}
\end{equation}

Within the semiclassical photodetection model, the detected photon number $K_b$ in each time bin $[t_b,t_b+\Delta t]$ is Poisson distributed conditional on the population trajectory~\cite{Mandel1964}:
\begin{equation}
	K_b\mid p_\uparrow
	\sim\operatorname{Poisson}\!\left[
	\eta\Gamma_1N
	\int_{t_b}^{t_b+\Delta t}p_\uparrow(t)\,\mathrm dt
	\right].
	\label{eq:photon_counts}
\end{equation}
The corresponding estimate of the bin-averaged population is
\begin{equation}
	\hat p_{\uparrow,b}
	=\frac{K_b}{\eta\Gamma_1N\Delta t}.
	\label{eq:population_estimate}
\end{equation}
Writing the population averaged over the bin as $\bar p_{\uparrow,b}$, the conditional counting variance is
\begin{equation}
	\operatorname{Var}\!\left(
	\hat p_{\uparrow,b}\mid p_\uparrow
	\right)
	=\frac{\bar p_{\uparrow,b}}
	{\eta\Gamma_1N\Delta t}.
	\label{eq:counting_variance}
\end{equation}
Thus, for a fixed population fraction, detection efficiency, and bin width, the standard deviation of the counting error decreases as $N^{-1/2}$.
Here the conditional Poisson sampling is a semiclassical readout model applied to the magnetization record, rather than a simulation of the exact measurement-conditioned density operator.

Figure~\ref{fig:photonreadout} compares the population trajectories and their photon-counting readout at $N=10^4$, $10^5$, and $10^6$, using the same initial preparation, detection efficiency $\eta=0.3$, and bin width $\Delta t=0.08T$.
To distinguish the spectral structure of the underlying dynamics from the additional counting background, we evaluate the power spectra of the bin-averaged population and its count estimate, denoted $S_{\mathrm{traj}}$ and $S_{\mathrm{count}}$.
Both spectra use the same $128T$ record following an $8T$ preparation interval and are computed with the Welch method~\cite{Welch1967}, using $32T$ Hann segments with $50\%$ overlap.
Their common normalization in absolute population units allows the oscillatory peaks and backgrounds to be compared directly.

The particle-number dependence reflects the combined effects of intrinsic finite-size fluctuations and photon-counting noise.
At $N=10^4$, intrinsic fluctuations strongly disrupt the regularity of the population dynamics, leaving no well-resolved oscillation peak even in the trajectory spectrum.
The absence of a clear spectral signature therefore originates in the underlying dynamics, rather than solely in the detection process.

At $N=10^5$, the weaker intrinsic fluctuations allow a more regular oscillatory signal to develop, but photon shot noise still largely obscures individual cycles in the time-domain record.
Spectral analysis nevertheless resolves a peak near $2\Omega$ approximately an order of magnitude above the surrounding background.
This distinction arises because the oscillatory power is concentrated near its characteristic frequency, whereas counting noise contributes a broadband background~\cite{Landi2024}.
The collective oscillation can therefore be identified spectrally even when its individual cycles are difficult to distinguish directly.

At $N=10^6$, the relative counting fluctuations are further suppressed, and the detected record closely follows the underlying population modulation.
The agreement between the two spectra around the dominant peak likewise shows that photon counting faithfully preserves the oscillatory signal.
For the detection conditions simulated here, shot noise thus has little effect on resolving the collective oscillations in either domain.

\subsection{Robustness to experimental parameter noise}
\label{sec:parameter_noise}

In addition to the intrinsic fluctuations considered above, noise in experimental control parameters can affect the stability of collective oscillations.
To assess their robustness against such noise, we introduce stochastic variations of the interaction strength $J$, local decay rate $\Gamma_1$, and pure-dephasing rate $\gamma_\phi$ into the mean-field Bloch equations.
We vary one parameter at a time, keeping the remaining independent parameters fixed.

\begin{figure}[tbp]
\centering
\includegraphics[width=\columnwidth]{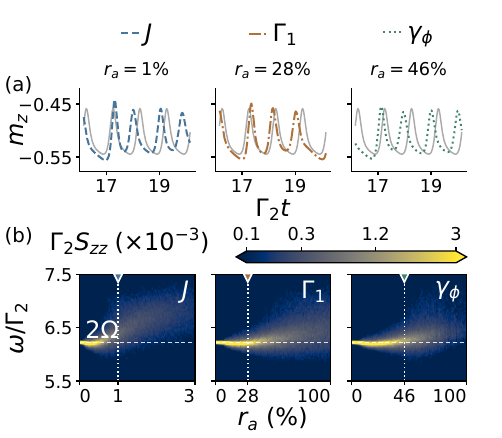}
\caption{Mean-field response to parameter noise.
    (a) Representative $m_z$ traces (colored) and noiseless references (gray).
    (b) Longitudinal power spectra $S_{zz}$ from $128T$ records, averaged over 100 realizations per strength.
    Vertical dotted lines and triangles mark the strengths in (a); horizontal dashed lines mark $2\Omega$.
    The common absolute-power color scale uses a square-root mapping.
    All realizations start from $\boldsymbol m(0)\simeq(0.10,0.01,-0.28)$, with the same system parameters as in Fig.~\ref{fig:semiclassical_switching}.}
\label{fig:parameter_noise}
\end{figure}

To keep the dissipative rates positive and preserve the ferromagnetic sign of the interaction, we use a common lognormal modulation,
\begin{equation}
	a(t)=a_0\exp\!\left[s_a u_a(t)-\frac{s_a^2}{2}\right],
	\qquad a\in\{J,\Gamma_1,\gamma_\phi\},
	\label{eq:positive_parameter_noise}
\end{equation}
where $a_0$ is the noiseless parameter value and $s_a$ is the logarithmic modulation amplitude.
The stationary Ornstein--Uhlenbeck process $u_a(t)$ has zero mean, unit variance, and correlation $\mathbb E[u_a(t)u_a(t')]=e^{-|t-t'|/\tau_c}$~\cite{Gillespie1996}.
The exponential form guarantees $a(t)>0$, while the offset $-s_a^2/2$ ensures $\mathbb E[a(t)]=a_0$.
Thus, the modulation introduces parameter fluctuations while preserving the mean parameter values.

For correlation times satisfying $\omega\tau_c\ll1$, the parameter-noise spectrum is approximately flat over the relevant frequency range.
We choose $\tau_c=0.01T$, where $T$ is the full period of the noiseless mean-field cycle, so that the fluctuations approximate white noise over the frequency range relevant to the collective oscillations.

The noise amplitude is specified by the relative standard deviation,
\begin{equation}
	r_a
	=\frac{\sqrt{\mathbb E[(a(t)-a_0)^2]}}{a_0}
	=\sqrt{e^{s_a^2}-1},
	\label{eq:relative_noise_strength}
\end{equation}
and is reported as a percentage.
At equal $r_a$ and $\tau_c$, the three channels have the same relative noise statistics, allowing their effects on the collective oscillations to be compared.
The transverse relaxation rate follows $\Gamma_2(t)=\Gamma_1(t)/2+2\gamma_\phi(t)$, while time and frequency are expressed using the noiseless value of $\Gamma_2$.
For each noise strength, $S_{zz}(\omega)$ in Fig.~\ref{fig:parameter_noise} is the average of the longitudinal magnetization power spectra from 100 independent realizations.

The different noise responses in Fig.~\ref{fig:parameter_noise} emphasize that robustness depends on which part of the feedback mechanism is perturbed.
These tests concern fluctuations around an operating point with nonzero dephasing; they complement, rather than replace, the necessity result obtained by removing dephasing altogether.

\section{Discussion and conclusions}
\label{sec:discussion}

Our results identify a route from a ubiquitous decoherence channel to autonomous temporal order in a familiar interacting spin model. Pure dephasing is energy neutral for isolated spins in the dephasing basis, but not for their transverse interaction energy: the loss of ferromagnetic correlations supplies energy that local decay subsequently removes. The key is not heating alone, but its delayed return as a collective feedback force. Equation~\eqref{eq:positive_feedback} makes this conversion explicit: the cycle-averaged feedback power is proportional to $\avg{\dot Q^2}$ and compensates damping, whereas instantaneous energy response produces no net work. The resulting design principle is to identify energy-bearing correlations, a dissipative channel that modifies them, and a relaxing internal variable that feeds back on collective motion. Unlike stabilization through reservoir memory~\cite{Das2026}, the delay here arises within dynamics generated by a time-independent Markovian master equation. This perspective suggests treating dephasing strength and basis as control parameters rather than simply minimizing decoherence. The exact amplitude--phase relation provides a diagnostic for pursuing this mechanism beyond fully connected ensembles, where spatial fluctuations can reshape temporal order~\cite{Daviet2025}. Metastability adds a complementary control problem: an oscillatory attractor can be repeatedly reactivated and can participate in transitions between stationary regions. For sustained readout and time-crystal sensing~\cite{CabotSensing2024,Gribben2025}, controlling access to that attractor can therefore matter alongside stabilizing its oscillation frequency.

We have shown that local pure dephasing can sustain continuous time-crystalline order in a spin-$1/2$ ensemble with local decay. We have established an exact energy identity, proved the necessity of dephasing for periodic motion in the homogeneous mean-field model, and identified delayed energy feedback as the mechanism that balances damping. We have connected this mechanism to quantum Liouvillian modes and collective correlations, uncovered recurrent oscillatory episodes within metastable switching, and characterized photon-counting readout and robustness. These results have established a constructive role for local coherence loss and identified a route to autonomous temporal order through a dynamical loop of dephasing-induced energy charging, decay-driven relaxation, and delayed feedback on the collective magnetization.

\begin{acknowledgments}
	This work was supported by the National Natural Science Foundation of China (Grants No.~12375013 and No.~12547109) and the Guangdong Provincial Quantum Science Strategic Initiative (Grant No.~GDZX2503008).
\end{acknowledgments}

\bibliographystyle{apsrev4-2}
\bibliography{references}

\end{document}